%% file: main_trg_paper.tex
\documentclass[review]{elsarticle}

\usepackage{standalone}
\usepackage{capt-of}
\usepackage{graphicx}
\usepackage{epstopdf}
\usepackage{subfigure}
\usepackage[utf8]{inputenc}
\usepackage[english]{babel}
\usepackage{amsmath}
\usepackage{hyperref}
\usepackage{rotating}
\usepackage{lineno}

\begin{document}\sloppy
\begin{frontmatter}

\title{LUX Trigger Efficiency}
\input{20180901-lux-tex-author-list_edited}

\begin{abstract}
The Large Underground Xenon experiment (LUX) searches for dark matter using a dual-phase xenon detector. LUX uses a custom-developed trigger system for event selection. In this paper, the trigger efficiency, which is defined as the probability that an event of interest is selected for offline analysis, is studied using raw data obtained from both electron recoil (ER) and nuclear recoil (NR) calibrations. The measured efficiency exceeds 98\% at a pulse area of 90 detected photons, which is well below the WIMP analysis threshold on the S2 pulse area. The efficiency also exceeds 98\% at recoil energies of \mbox{0.2 keV} and above for ER, and \mbox{1.3 keV} and above for NR. The measured trigger efficiency varies between 99\% and 100\% over the fiducial volume of the detector.
\end{abstract}

\begin{keyword}
Trigger efficiency \sep Dark matter detectors \sep WIMPs \sep Liquid xenon
\end{keyword}

\end{frontmatter}

\section{The LUX Trigger} \label{LUXIntro}
The Large Underground Xenon experiment (LUX) searches for dark matter particles using a dual-phase xenon detector, described in detail in Ref. \cite{LUX_NIMPaper_2013}. The primary goal of LUX is to search for Weakly Interacting Massive Particles (WIMPs). An energy deposition from an incident particle produces prompt scintillation light, called S1, and ionization electrons. The ionization electrons drift to the liquid surface and are extracted into the gas region by electric fields to produce electroluminescence light, called S2. The S1 and S2 signals are detected by two photomultiplier tube (PMT) arrays, located at the top and bottom of the detector. Each array consists of 61 Hamamatsu R8778 PMTs. Further details on the PMTs can be found in Ref. \cite{LUX_PMTPaper}. 

\begin{figure*}[ht]
	\centering
    \includegraphics[width=1.0\linewidth]{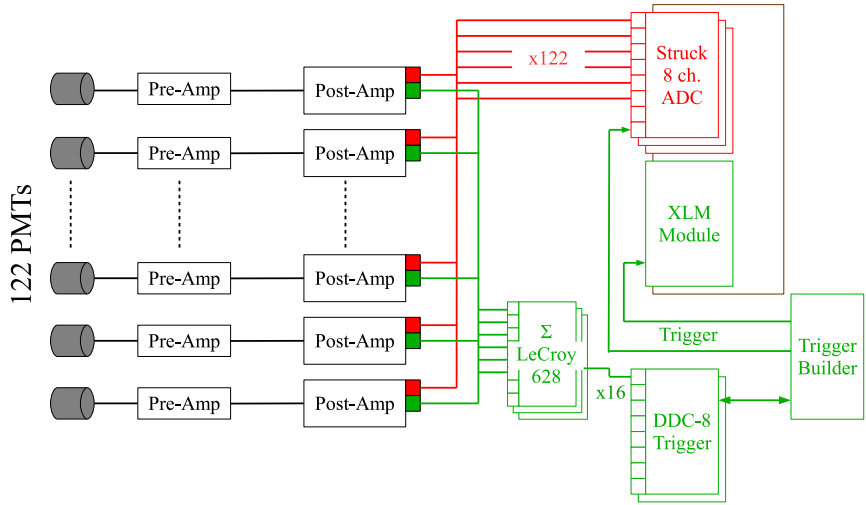}
    \captionof{figure}{Diagram of the LUX electronics. The LUX trigger system consists of three separate modules, i.e. the DDC-8, the trigger builder, and the XLM modules.}
	\label{Fig:LUXDAQ_schematic}
\end{figure*}

\begin{figure}[ht]
	\centering
    \includegraphics[width=1.0\linewidth]{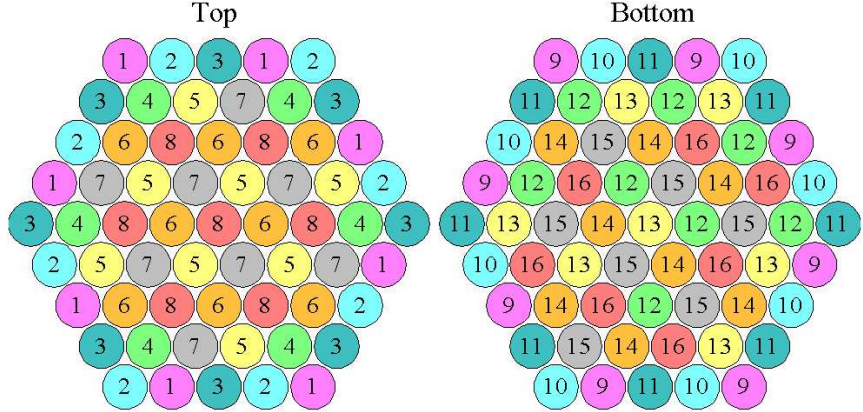}
    \captionof{figure}{Trigger group summing configurations of the top and bottom PMT arrays. Each circle represents each PMT, and the number inside indicates the assigned trigger group. The difference between the top and bottom patterns is for the maximum-based trigger. The maximum-based trigger is discussed in Ref. \cite{LUX_TRG_Paper}, and are not used in LUX WIMP searches.}
	\label{Fig:TrgSumConfig}
\end{figure}

A diagram of the LUX electronics is shown in Fig. \ref{Fig:LUXDAQ_schematic}. The PMT signals are first amplified by pre-amplifiers that are installed as close as possible to the PMTs to minimize the cable capacitance, and thus noise. Post-amplifiers are responsible for pulse shaping and amplification to match the requirements of the digitizers of the data acquisition (DAQ) and the LUX trigger systems, such as the sampling frequency. All individual PMT signals are digitized by the DAQ digitizers without requiring the presence of trigger signals and recorded onto dedicated data storage disks \cite{LUX_DAQ_2012}. In a separate trigger chain, the PMT signals are summed into 16 trigger groups, using the patterns shown in Fig. \ref{Fig:TrgSumConfig}. Each group contains 6-8 PMT signals. The summing configuration is chosen such that no neighboring PMTs belong to the same trigger group to minimize signal saturation during a large energy deposition event. The trigger system processes these signals to identify the periods in which pulses of interest occur. Trigger signals from the trigger system are recorded along with the PMT signals, digitized by the DAQ digitizers. The waveforms in the trigger chain, from which the trigger signals are obtained, are not recorded. An event is defined by a time window of \mbox{0.5 ms} before and after a trigger signal. A detailed discussion of the LUX DAQ system and the LUX trigger system can be found in Refs. \cite{LUX_DAQ_2012} and \cite{LUX_TRG_Paper}.

\begin{figure}[ht]
	\centering
    \includegraphics[width=88mm]{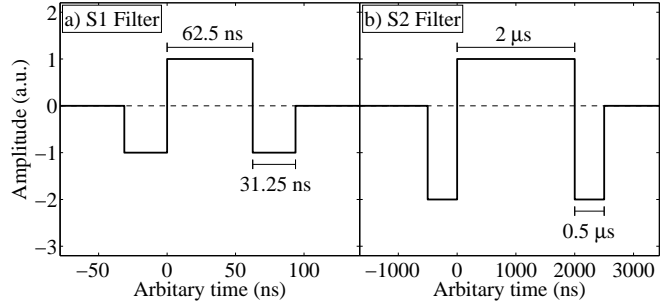}
    \captionof{figure}{Depictions of the S1 and S2 filters. The central lobes of both filters are responsible for pulse area running integration and the two side lobes are responsible for baseline subtraction. The length of the central lobe, the so-called filter width, is designed to be similar to the respective pulse type; \mbox{62.5 ns} for the S1 filter and \mbox{2 $\mu$s} for the S2 filter. The pulse area corresponds to the value of the filter output when the entire pulse is enclosed within the central lobe.}
	\label{Fig:S1FS2F}
\end{figure}

The LUX trigger system is responsible for event selection, based on pulse shape characteristics, utilizing computing power provided by the Spartan-6 FPGA. The pulse area in each trigger group is calculated using two digital filters, called the S1 and S2 filters, that perform baseline subtraction and pulse area running integration. Depictions of the S1 and S2 filters are shown in \mbox{Fig. \ref{Fig:S1FS2F}}. In this work, pulse areas are presented in units of detected photons (phd) which accounts for the possibility that two photoelectrons are created at the PMT photocathode by one incident photon. Further discussion on this effect can be found in Ref. \cite{LUX_Run3_Reanalysis, DpheEmission}. The S1 and S2 filters are designed to select events based on the presence of S1 and S2 signals, respectively. The S1 filter is used when the S1 is the dominant signal, such as when operating without electric fields. The S2 filter is used when S2 signals are expected as part of events of interest. Both filters can also be used at the same time to select events consisting of S1 and S2 signals. During LUX WIMP searches, the total event rate in the detector was sufficiently low to ensure minimum dead-times of both DAQ and trigger systems. Also, to maintain sensitivity to small energy depositions that could produce very small S1 signals, only the S2 filter was implemented in the trigger decision.

\begin{figure*}[ht]
	\centering
    \includegraphics[width=1.0\linewidth]{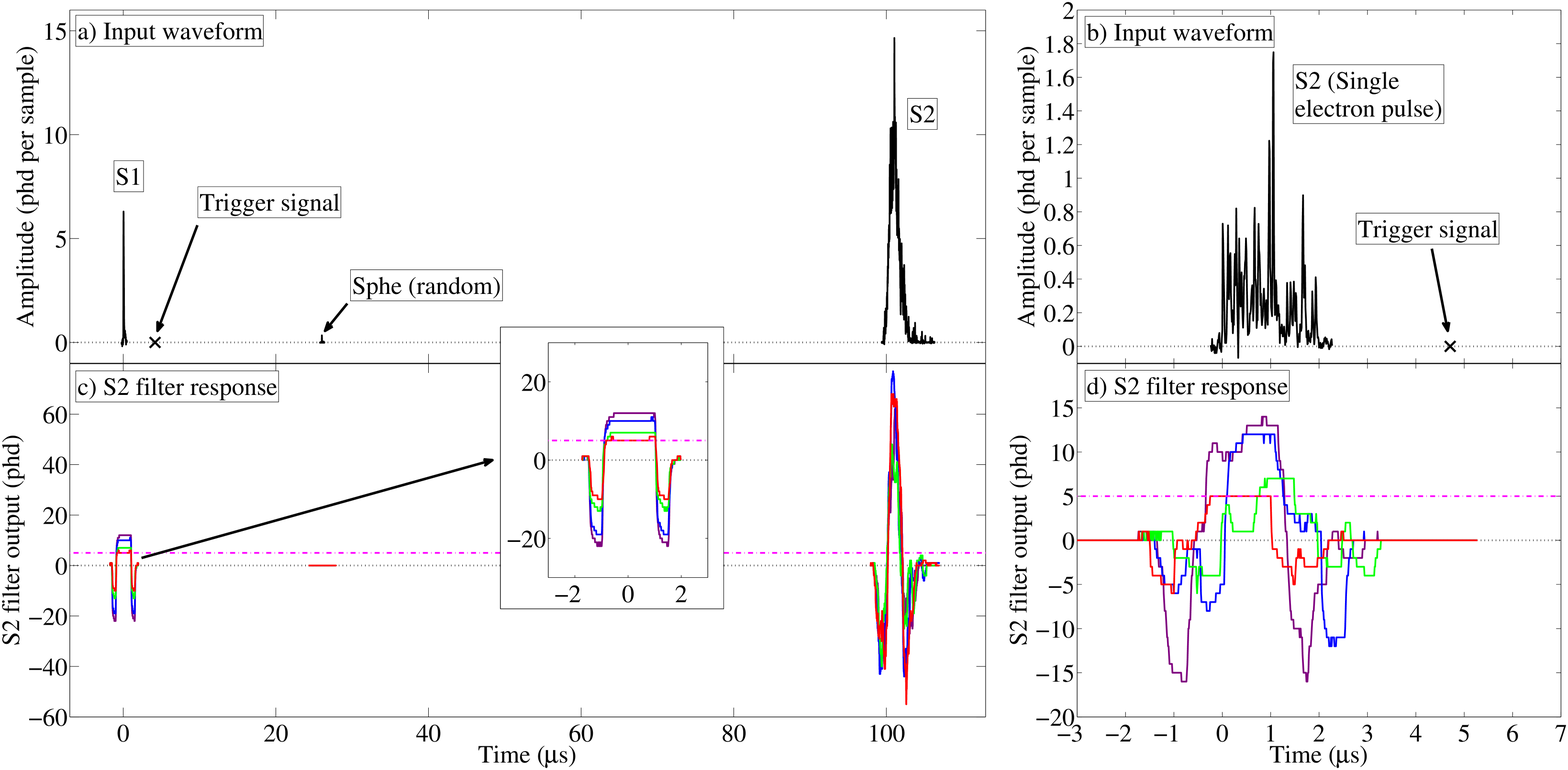}
    \captionof{figure}{S2 filter responses of example S1 and S2 signals. a) An input waveform of an event consists of an S1 and an S2 signal with a trigger signal denoted with a cross. The trigger signal is generated by the S1 signal. The small feature between the S1 and the S2 signals at around \mbox{27 $\mu$s} is a random single photoelectron pulse (sphe). b) An input waveform corresponding to an S2 signal associated with a single electron extracted from the liquid xenon surface. The waveforms in a) and b) are from the DAQ digitizers, which digitize waveforms at \mbox{100 MHz} sampling frequency. The amplitude is calibrated in units of phd per sample, such that the integral of the waveforms yields pulse areas in phd. The S2 filter outputs from the four trigger groups with the four largest pulse areas associated with the pulse that generates the trigger of the input waveform in a) and b) are shown in c) and d), respectively, demonstrating typical S2 filter responses of S1 and S2 signals. In both examples, at least two trigger groups have the S2 filter outputs greater than the low filter threshold which is represented with the horizontal dash-dotted line (color available online).}
	\label{Fig:S2FResponse}
\end{figure*}

The S2 filter can respond to both S1 and S2 signals, as demonstrated in Fig. \ref{Fig:S2FResponse}. Two thresholds can be applied to the output of the S2 filter. The first threshold is called the low filter threshold and the second threshold is called the veto threshold. A valid event requires a filter output above the low filter threshold and below the veto threshold. During WIMP searches, the low filter threshold was set at approximately \mbox{5 phd}, equivalent to approximately $\frac{1}{5}$ of the measured average pulse area of a signal associated with a single electron extracted from the liquid xenon surface \cite{LUX_Run3_Reanalysis}. The veto threshold was disabled to maintain sensitivity to high energy events. The S2 filter width was set at \mbox{2 $\mu$s}, which was the anticipated largest S2 pulse width. Throughout this work, the pulse width is defined as the time difference between the times at which the cumulative pulse area of the pulse crosses 2\% and 98\% of the total pulse area. These crossing times define the pulse start and the pulse end times used in this work. In LUX, S2 signals that are wider than \mbox{2 $\mu$s} are observed. However, the S2 filter width is kept at \mbox{2 $\mu$s} to minimize baseline noise pick-up. The effect of the S2 pulse width being larger than the S2 filter width is discussed in section \ref{subsection:TrgEff_vs_width}. An event of interest was required to have at least two trigger groups with S2 filter outputs greater than the low filter threshold within a time window, called the trigger search window, which was set at \mbox{2 $\mu$s}. If pulses meeting these requirements are found, a trigger signal is sent to the DAQ and the trigger system is put in an inactive mode for a user-defined time period, called the hold-off time. The hold-off time is used to ensure that pulses that are part of the same event do not generate multiple triggers and to prevent data overflow during a large energy deposition event. Prior to May 23\textsuperscript{rd}, 2013, the hold-off time was conservatively set to \mbox{4 ms}. It was reduced to \mbox{1 ms} on May 23\textsuperscript{rd}, 2013, to maximize the acquisition live-time \cite{LUX_TRG_Paper}. This change increases the trigger system live-time from 96\% to 99\% and has been verified to have no negative impact on the data quality. 

LUX was operated at the Sanford Underground Research Facility (SURF) in Lead, SD, between October 2012 and October 2016. Dark matter search results from LUX can be found in Refs. \cite{LUX_Run3_Reanalysis,LUX_FirstRunResults_2014,LUX_Run3_plus_4,LUX_Spin_Dep_1,LUX_Spin_Dep_2,Axion_Result}.

In this paper, the trigger efficiency measurement is described in Section \ref{section:Technique_Measure_TrgEff}. The dependence of the trigger efficiency on various pulse shape characteristics is discussed in Section \ref{section:PulseBased_TrgEff}. The dependence of the trigger efficiency on event properties is discussed in Section \ref{section:EventBased_TrgEff}. The paper is summarized in Section \ref{section:summary}.

\section{Measuring Trigger Efficiency} \label{section:Technique_Measure_TrgEff}
The trigger efficiency is an important parameter required for the interpretation of dark matter search results. The trigger efficiency is defined as the probability that an event of interest is selected for offline analysis. In this work, the trigger efficiency associated with the LUX WIMP analysis is presented. The efficiency is measured and studied as a function of various parameters to ensure a full understanding of the LUX trigger system. Throughout this work, raw data files collected by the LUX DAQ system, running in the same mode as used during LUX WIMP searches, are used to determine the trigger efficiency. 

\begin{figure} [ht]
    \centering
    \includegraphics[width=88mm]{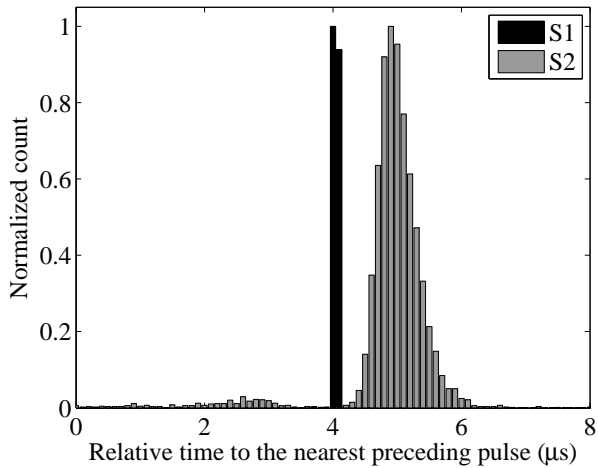}
    \captionof{figure}{Time difference between the trigger signal and pulse start time of the nearest preceding pulse. The histogram uses a bin size of \mbox{0.1 $\mu$s}. The height of the peaks of the distributions are normalized to one. Only a subset of the ER data that is used in this work is included in this analysis. For a signal that generates a trigger, the trigger signal occurs at the end of the trigger decision process, which takes a specific time after the process is started. For S1 signals which have a short rise time, the trigger signal latency times are clustered together because the associated S2 filter outputs promptly cross the threshold and start the trigger decision process. For S2 signals which have a longer rise time than S1 signals, the associated S2 filter outputs may cross the threshold at different times after the pulse start time. The trigger decision process of these signals do not always start at the pulse start time.}
    \label{Fig:TrgPulseTimeDist}
\end{figure}

The waveforms from each PMT are reconstructed from the data stored in the raw data files and then processed with the same data analysis framework as is used in the LUX WIMP analysis. In this work, each entire data set is processed, independently of the presence of trigger signals, while only the time windows around the trigger signals are processed in the LUX WIMP analysis. The analysis framework performs waveform analysis to locate and extract the properties of each pulse, such as pulse type (e.g., S1 or S2), pulse area, pulse timing, and pulse width. In addition, for each pulse found, it is determined whether that pulse generated a trigger, based on its timing and the presence of a nearby trigger signal. The trigger system takes a well-defined time to make a trigger decision once an S2 filter output crosses the low filter threshold, and a trigger signal is generated at the end of the trigger decision process. The trigger latency time is measured to be \mbox{4 $\mu$s}, as shown in Fig. \ref{Fig:TrgPulseTimeDist}. Due to the short rise time of the S1 signals, the associated S2 filter outputs promptly cross the threshold and start the trigger decision process. For these signals, the latency time of the trigger signal is well-defined. For S2 signals, which have a longer rise time than S1 signals, the S2 filter outputs can cross the threshold much later than the pulse start time, and the latency time covers a larger time window. A pulse is considered as having generated a trigger if there is a trigger signal within a time window of \mbox{4 $\mu$s} after the pulse start time to \mbox{4 $\mu$s} after the pulse end time. 

A population of small S2 signals, associated with single electron pulses (SEs), is observed which have relative times to the trigger signals between 2.0 and \mbox{3.0 $\mu$s}. These SEs have a strong temporal correlation with the trigger signals, but they cannot be the pulses that generate the triggers, based on the trigger decision process \cite{LUX_TRG_Paper}. Each of the SEs follows a large S1 signal. The time between the S1 signals and the trigger signals is found to be \mbox{4 $\mu$s}, which implies that the trigger signals are generated by the S1 signals. The time between these SEs and the preceding large S1 signals is 1.0-\mbox{2.0 $\mu$s}, suggesting that these electrons originate from the region between 2 and \mbox{6 mm} below the liquid xenon surface. This region is consistent with the position of the high voltage grid (gate grid) which is located at about 5 mm below the liquid xenon surface \cite{LUX_NIMPaper_2013}. In this region, the electric field strength is \mbox{3.1 kV/cm} \cite{LUX_FirstRunResults_2014}, and the electron drift velocity in liquid xenon at this electric field is about 2-\mbox{3 mm/$\mu$s} \cite{ElectronDriftVelocity}. These observations suggest that these electrons are very likely ionized from the high voltage grid by the S1 photons. Pulses of this character are therefore removed from the subsequent analysis.

The LUX trigger system makes the trigger decision based on individual pulses. The efficiency of this process is called the pulse-based efficiency. Many important WIMP-search parameters require both S1 and S2 signals to be present and considered together. The trigger efficiency evaluated based on a combination of S1 and S2 signals is called the event-based trigger efficiency. In both cases, the trigger efficiency is defined as, 

\begin{equation} \label{equation:pulse_based_TrgEff_Def}
\mathcal{E}(A) = \frac{N_{Trg}(A)}{N_{Tot}(A)} \times 100 \%,
\end{equation}

\noindent where $A$ is the property of interest, $\mathcal{E}(A)$ is the pulse-based(event-based) trigger efficiency for pulses(events) with properties $A$, $N_{Trg}(A)$ is the number of pulses(events) that generate triggers with properties $A$, and $N_{Tot}(A)$ is the number of all pulses(events) with properties $A$. The properties of interest can be a single property, such as pulse area, pulse width, and recoil energy, or a combination, such as having pulse areas of 50-\mbox{60 phd} and pulse widths of 0.5-\mbox{1.0 $\mu$s}. The trigger efficiency measurement is limited by the efficiencies of the LUX analysis framework involved in performing its function. For example, if the analysis framework fails to recognize a pulse, that pulse will not be accounted for in the trigger efficiency measurement. Similarly, if a characteristic of a pulse is miscalculated due to limitations, such as having a small pulse area, that pulse will contribute incorrectly to the measurement. The impact of the above factors on the overall performance and efficiency of the LUX analysis framework have been thoroughly examined \cite{LUX_PRD}, and are based on both ER and NR calibration data. The measurement of the trigger efficiency utilizes a subset of both ER and NR calibration data, tritiated methane (CH\textsubscript{3}T), and D-D (DD), respectively \cite{LUX_CH3T_Calib, LUX_DD_Paper}, collected between November and December of 2013. During these calibrations, the trigger settings were identical to the trigger setting used during the LUX WIMP searches. A benchmark of 98\% for the trigger efficiency is chosen such that the impact of missing triggers is a subdominant factor in the overall system inefficiency. A measurement shows that the baseline fluctuation crosses the S2 low filter threshold at a rate of about \mbox{$10^{-6}$ Hz} for each trigger group. Therefore, the contribution from baseline noise is expected to be subdominant when the efficiency has reached 98\%, and is not explicitly studied further. A binomial confidence interval at 95\% confidence level is used to represent the uncertainty associated with the efficiency. The trigger system is not saturated due to high event rates during WIMP searches and calibrations. Event rate saturation in the trigger system is dominated by the hold-off time, corresponding to rate limits reaching 50\% dead time at \mbox{1 kHz} for a \mbox{1 ms} hold-off time and \mbox{125 Hz} for \mbox{4 ms}. Average event rates during WIMP searches were \mbox{9 Hz}, when a \mbox{4 ms} hold-off was used, and \mbox{12 Hz}, when a \mbox{1 ms} hold-off was used. The increase in the trigger rate is mostly due to the delayed electrons emission following large S2 signals \cite{LUX_Run3_Reanalysis}. Peak event rates during calibrations, where a \mbox{1 ms} hold-off was used, were less than \mbox{100 Hz}. Thus, in all cases, the event rates are significantly lower than the limit from the hold-off time. When the hold-off time is disabled, the trigger system is capable of handling more than \mbox{1 kHz} of event rate without significant dead time \cite{Eryk_Thesis}.

\section{Pulse-based Trigger Efficiency} \label{section:PulseBased_TrgEff}

The LUX trigger system responds differently to pulses with different pulse shape characteristics. In this section, the effects of pulse area and pulse width on the trigger efficiency are discussed. S1 and S2 signals are studied separately due to differences in their pulse shapes. 

\subsection{Dependence on Pulse Area} \label{subsection:TrgEff_vs_RawTotalArea}

Many detector-dependent effects impact the measurement of the S1 and the S2 pulse areas. Examples are the probability that an S1 photon is detected by the PMTs, called the light collection efficiency, electron losses due to impurities in the liquid xenon, and non-uniform extraction fields. Consequently, a precise measurement of the energy deposition requires pulse area corrections. The average pulse area correction is typically a few percent for S1s, while for S2s, due to the strong dependence on liquid xenon purity, the correction is approximately 20-30\% \cite{AttilaDobi_thesis_2014_S2S2Correction,RichardKnoche_thesis_2014_S2S2Correction}. Since the LUX trigger system makes trigger decisions based on the uncorrected signals, this study uses the pulse areas found by the analysis framework without any detector-dependent corrections.

\begin{figure}[ht]
\centering
    \includegraphics[width=88mm]{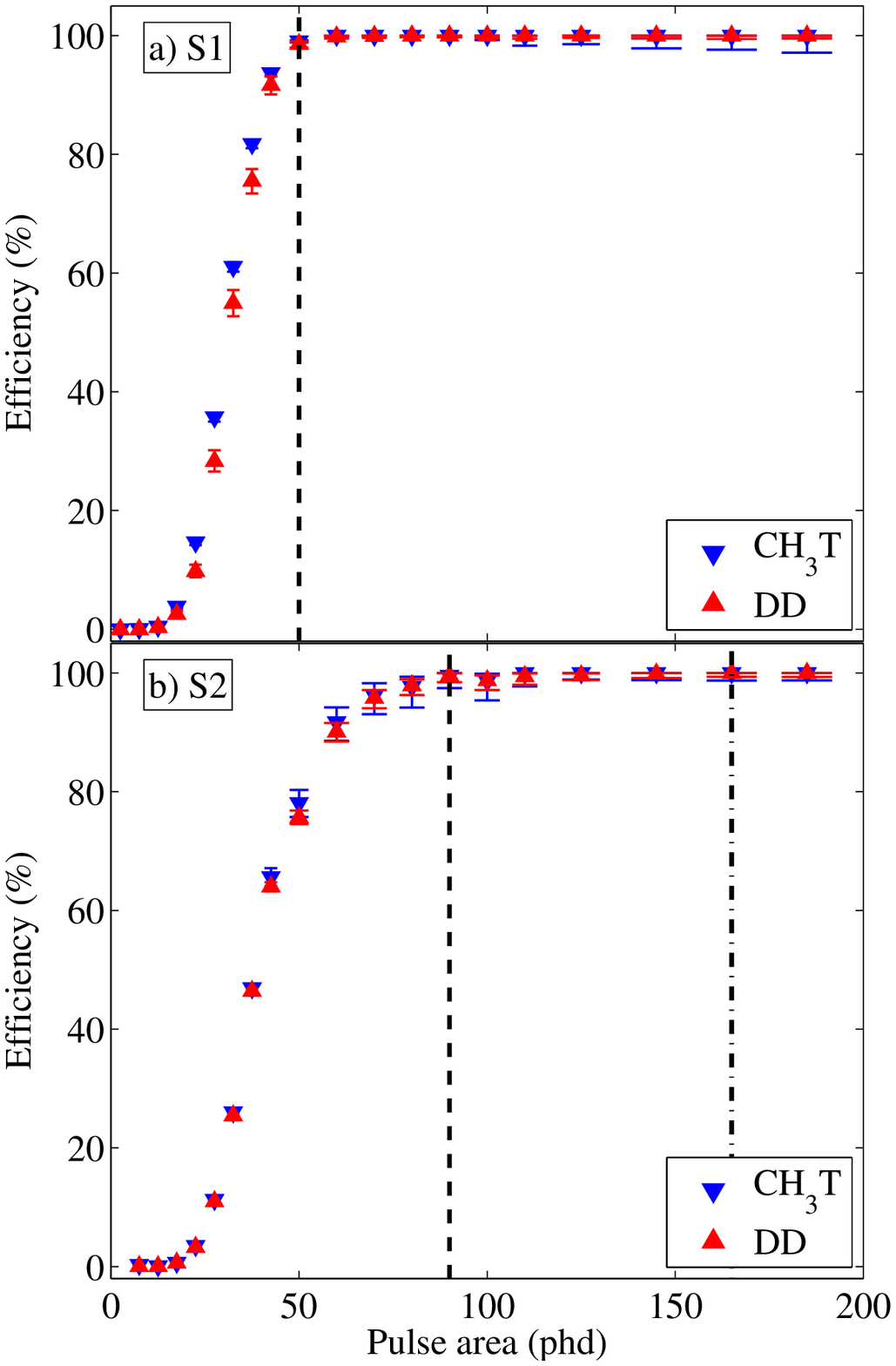}
    \captionof{figure}{Measured trigger efficiency as a function of total pulse area. The results from CH\textsubscript{3}T and DD data are compared. a) The S2 filter trigger efficiency for S1 ($\mathcal{E}_{S1}(phd)$). b) The S2 filter trigger efficiency for S2 ($\mathcal{E}_{S2}(phd)$). The vertical dashed lines represent the pulse areas at which $\mathcal{E}_{S1}(phd)$ and $\mathcal{E}_{S2}(phd)$ exceed 98\%. They are at \mbox{50 phd} and \mbox{90 phd}, respectively. The dash-dotted line represents the lowest analysis threshold on the S2 signal used in the LUX WIMP analysis which is \mbox{165 phd}.}
    \label{Fig:TrgEff_vs_TotalArea}
\end{figure}

The measured trigger efficiency, as a function of pulse area, is shown in \mbox{Fig. \ref{Fig:TrgEff_vs_TotalArea}}. This figure shows that the S2 filter trigger efficiency for S1 ($\mathcal{E}_{S1}(phd)$) is greater than 98\% when the S1 pulse area is larger than \mbox{50 phd}. The S2 filter trigger efficiency for S2 ($\mathcal{E}_{S2}(phd)$) exceeds 98\% for S2 pulse areas larger than \mbox{90 phd}, which is equivalent to approximately 4 extracted electrons from the liquid xenon surface. The consistency of efficiencies measured from CH\textsubscript{3}T and DD data for S1 and S2 trigger indicates that the LUX trigger system performs its functions in S2 search mode essentially independently of the recoil types.

\begin{figure}[ht]
	\centering
    \includegraphics[width=88mm]{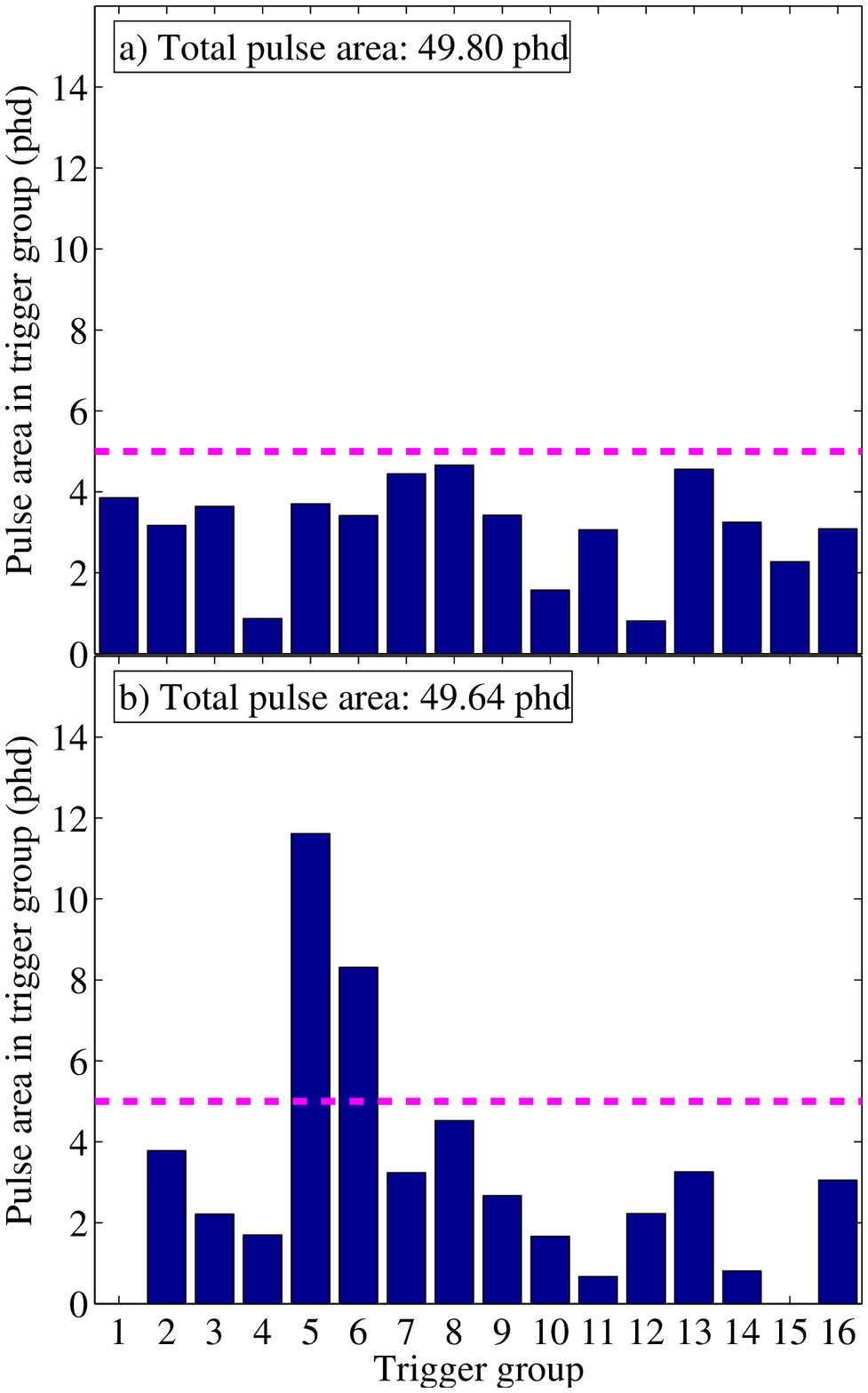}
    \captionof{figure}{Examples of two small S2 signals of similar size demonstrating the dependence of the trigger decision on the pulse area distribution among the 16 trigger groups. The low threshold of \mbox{5 phd} is denoted with the horizontal dashed lines. a) None of the trigger groups have a pulse area greater than the threshold; this pulse does not generate a trigger. b) Two trigger groups have pulse areas greater than the threshold; this pulse generates a trigger.}
    \label{Fig:SE_trg_group_area_dist}
\end{figure}

$\mathcal{E}_{S1}(phd)$ starts to rise from zero toward 100\% efficiency at lower pulse area than $\mathcal{E}_{S2}(phd)$. For instance, $\mathcal{E}_{S1}(phd)$ reaches 50\% at an S1 pulse area of approximately \mbox{30 phd}, while $\mathcal{E}_{S2}(phd)$ reaches that point at an S2 pulse area of approximately \mbox{40 phd}. Both 50\% efficiency points are greater than \mbox{10 phd}, which is twice the threshold requirement used per trigger group. Since the coincidence requirement is used, the trigger decision does not only depend on the total pulse area, but also on the pulse area distribution among the trigger groups. Two S2s of similar pulse area but different pulse area distribution are shown in Fig. \ref{Fig:SE_trg_group_area_dist} to demonstrate this dependency. For a pulse to have a pulse area in two trigger groups greater than \mbox{5 phd} on average, its total pulse area has to be much larger than \mbox{10 phd}. The difference of pulse areas at which the efficiencies exceed 50\% is not understood. Further study of this behavior, discussed in Section \ref{subsection:TrgEff_vs_width}, rules out the width of the S2 signal being longer than the S2 filter width as the cause. This pulse area range is well below the WIMP analysis threshold on the S2 pulse area of \mbox{165 phd} \cite{LUX_Run3_Reanalysis}, and does not effect the LUX WIMP analysis.

\subsection{Dependence on the Second Largest Pulse Area} \label{subsection:TrgEff_vs_second_area}
The LUX trigger system generates a trigger when at least two trigger groups have an S2 filter output greater than the \mbox{5 phd} threshold. In other words, a trigger will be generated when the two largest pulse areas among the 16 trigger groups are both greater than \mbox{5 phd}. Thus, the parameter that drives the trigger decision is the pulse area in the trigger group that has the second largest pulse area among the 16 trigger groups (SLPA). Studying the trigger efficiency based on this parameter shows the influence of the coincidence requirement on the trigger efficiency.

\begin{figure}[ht]
    \centering
    \includegraphics[width=88mm]{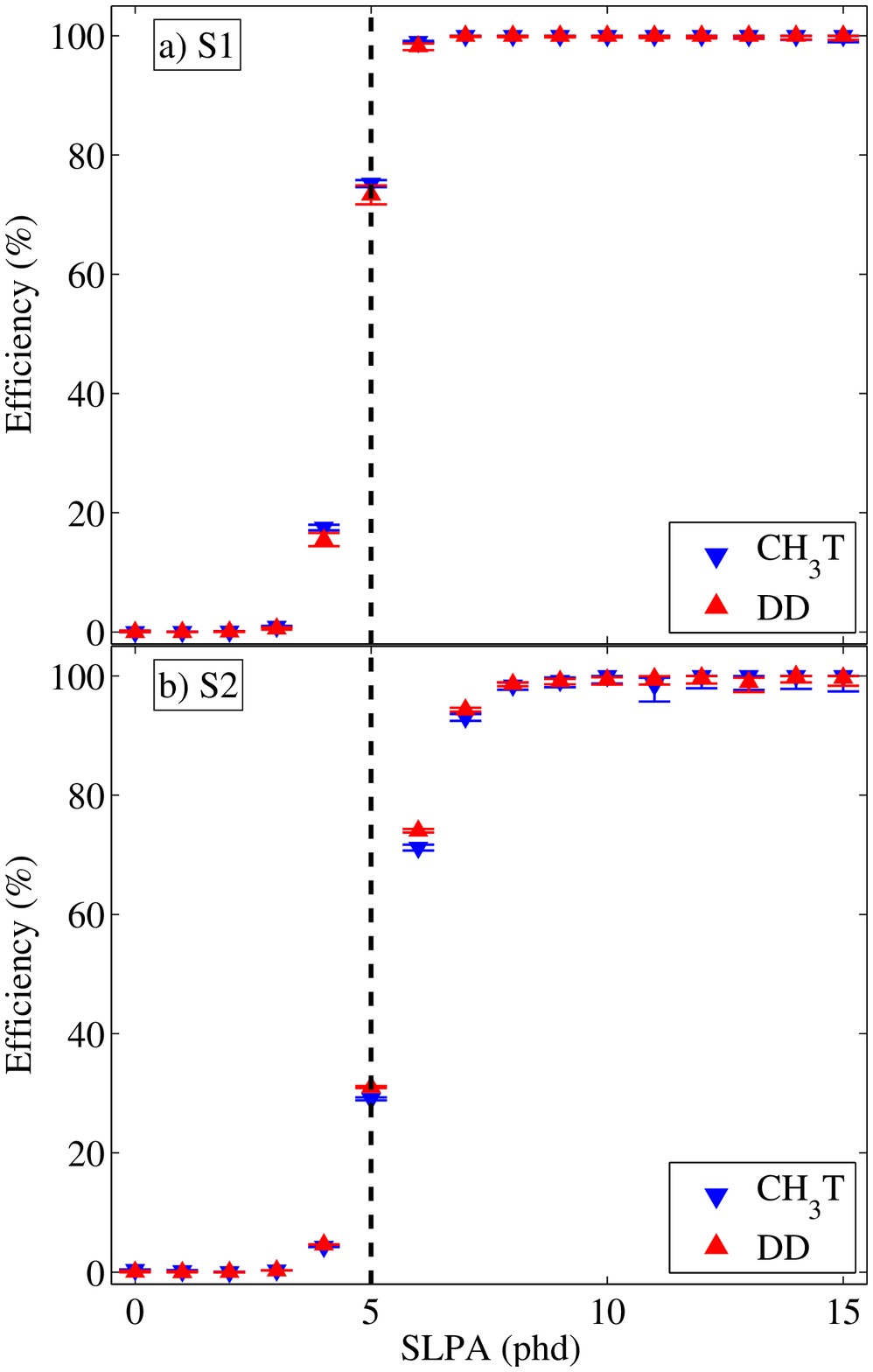}
    \captionof{figure}{Measured trigger efficiencies as a function of the SLPA. The results from CH\textsubscript{3}T and DD data sets are compared. a) The S2 filter trigger efficiency for S1 ($\mathcal{E}_{S1}(SLPA)$). b) The S2 filter trigger efficiency for S2 ($\mathcal{E}_{S2}(SLPA)$). The vertical dashed lines represent the low filter threshold of \mbox{5 phd}.}
    \label{Fig:TrgEff_vs_SecondArea}
\end{figure}

The trigger efficiency as a function of the SLPA is shown in Fig. \ref{Fig:TrgEff_vs_SecondArea}, and demonstrates that the efficiencies exceed 98\% for SLPA values above \mbox{7 phd} for S1 signals and 8 phd for S2 signals. Both $\mathcal{E}_{S1}(SLPA)$ and $\mathcal{E}_{S2}(SLPA)$ are not zero for SLPA values less than \mbox{5 phd} which is the threshold. The baseline noise contribution from the LeCroy 628 Linear Fan-in/Fan-out used in the trigger chain and the overall differences between the waveforms in the DAQ digitizers and the trigger chains are expected to be the causes of this behavior. 

$\mathcal{E}_{S1}(SLPA)$ starts to rise from zero toward 100\% at a lower SLPA value than $\mathcal{E}_{S2}(SLPA)$, a behavior that is similar to that seen in Fig. \ref{Fig:TrgEff_vs_TotalArea}. In this parameter space, $\mathcal{E}_{S1}(SLPA)$ and $\mathcal{E}_{S2}(SLPA)$ reach 50\% at an SLPA value of approximately 4.5 and \mbox{5.5 phd}, respectively. Since the exact waveforms processed by the LUX trigger system are not preserved, further study of this behavior is not possible.

\subsection{Dependence on Pulse Width} \label{subsection:TrgEff_vs_width}
The pulse area integration region in the trigger system is defined by the filter width. When an input pulse is wider than the S2 filter width, the pulse area reported by the S2 filter ($Area_{S2F}$) will be less than the actual pulse area of the input pulse returned by the LUX analysis framework \cite{LUX_TRG_Paper}. In LUX, the dispersion of the ionization electrons as they drift through the liquid xenon \cite{Sorensen_2011} to the liquid xenon surface can cause the S2 width to be as large as \mbox{4 $\mu$s}, which corresponds to an S2 signal of an event near the bottom of the detector \cite{LUX_DAQ_2012}. Increasing the S2 filter width also increases the baseline noise pick-up. An analysis shows that the low filter threshold would need to be increased significantly to suppress the noise pick-up, and thus reducing the efficiency for low-energy recoils \cite{Eryk_Thesis}. Therefore, the S2 filter width was kept at \mbox{2 $\mu$s}.

\begin{figure}[ht] 
    \centering
    \includegraphics[width=88mm]{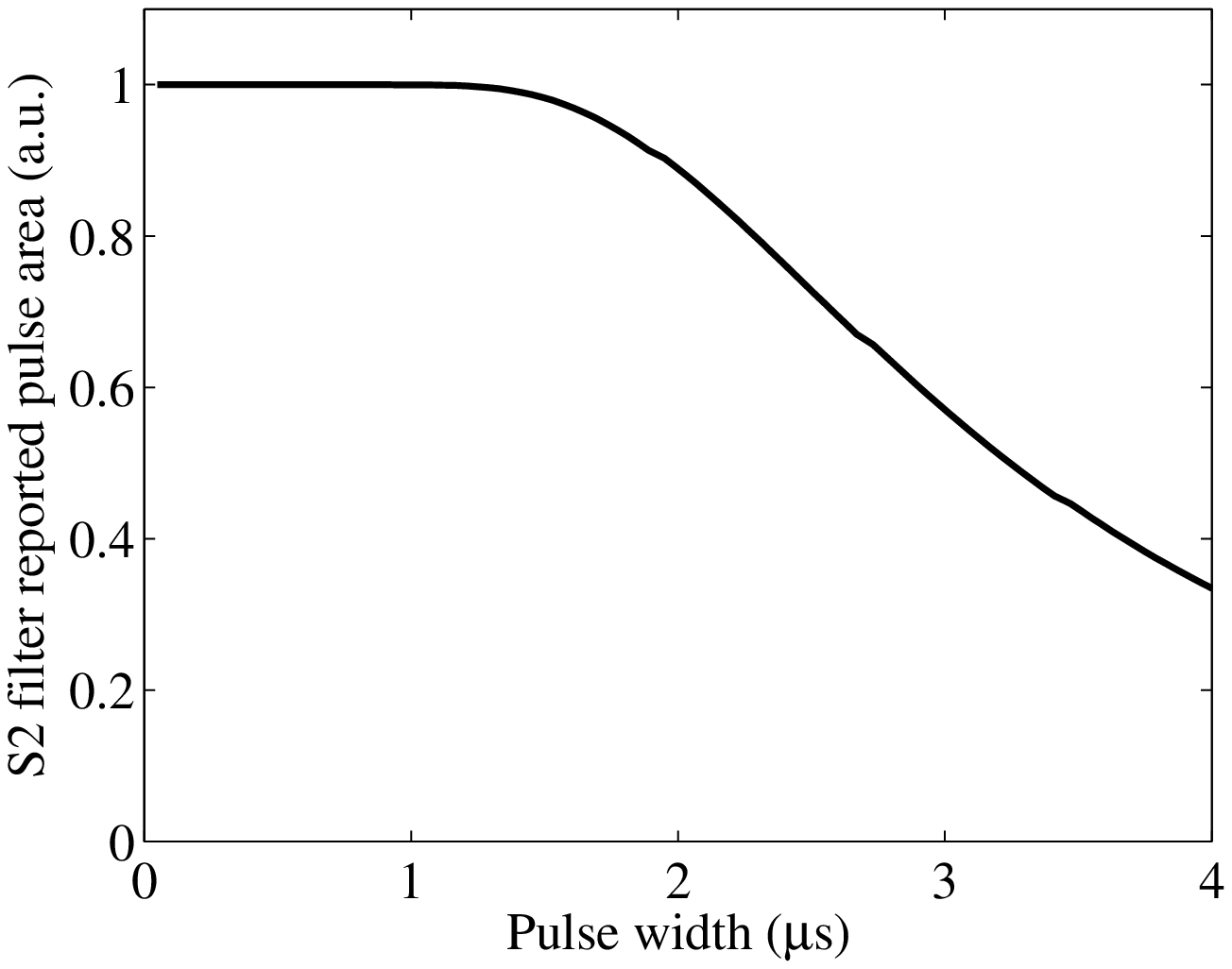}
    \captionof{figure}{Simulation results of the $Area_{S2F}$ of a Gaussian-shaped pulse of area one when the width of the input pulse is varied.}
    \label{Fig:S2F_MeasuredPulseArea_Width}
\end{figure}

To study the dependence of the S2 filter response on pulse width, a simulation has been carried out. Figure \ref{Fig:S2F_MeasuredPulseArea_Width} shows the relation between the $Area_{S2F}$ of Gaussian-shaped pulses of area one and the input pulse width, defined as the time difference between the pulse start and pulse end times. We observe that the $Area_{S2F}$ becomes smaller than the actual pulse area when the pulse width is comparable to the S2 filter width.

\begin{figure}[ht]
    \centering
    \includegraphics[width=88mm]{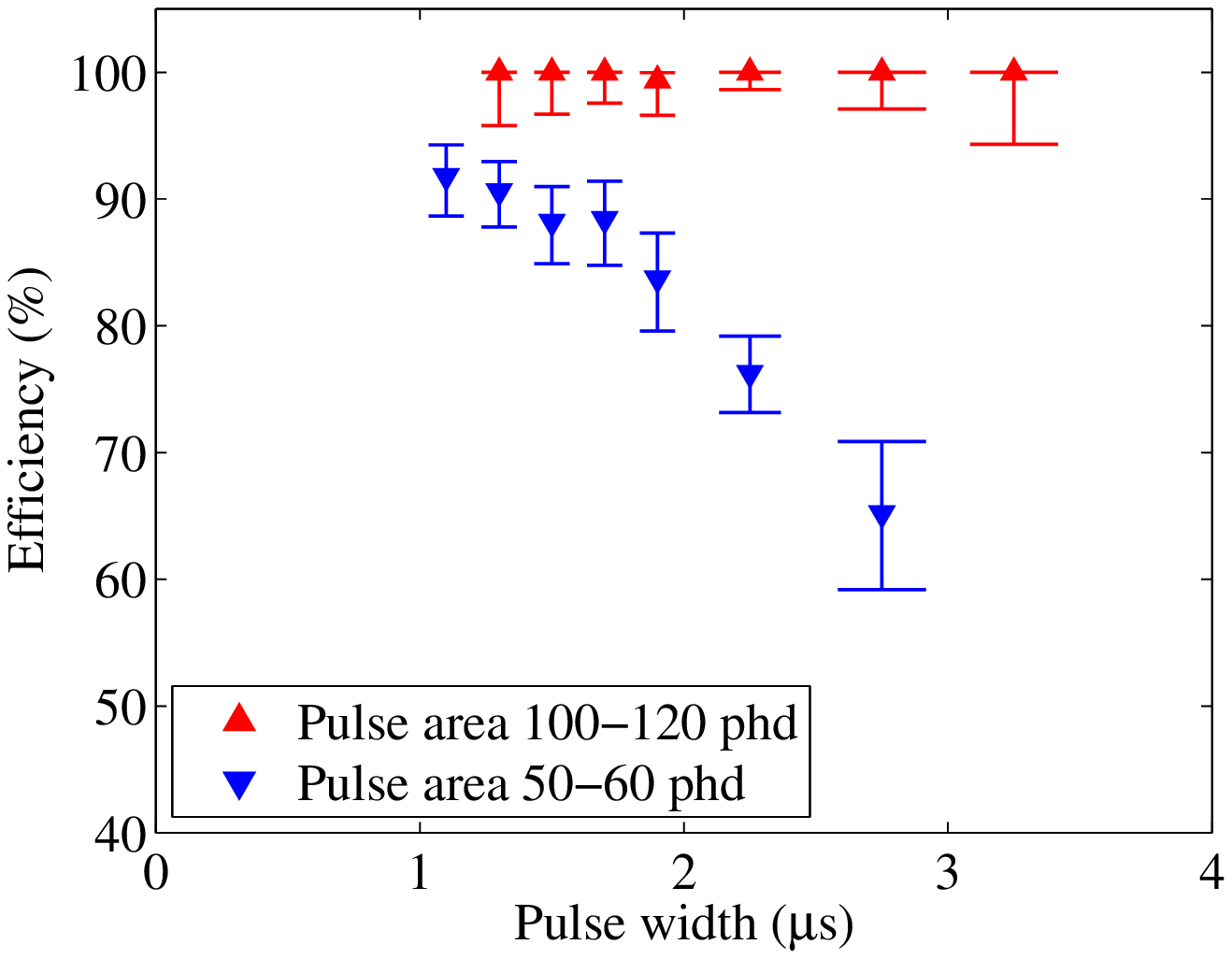}
    \captionof{figure}{Trigger efficiency as a function of S2 pulse width for two different pulse area ranges.}
    \label{Fig:TrgEff_vs_Width}
\end{figure}

\begin{figure}[ht]
    \centering
    \includegraphics[width=88mm]{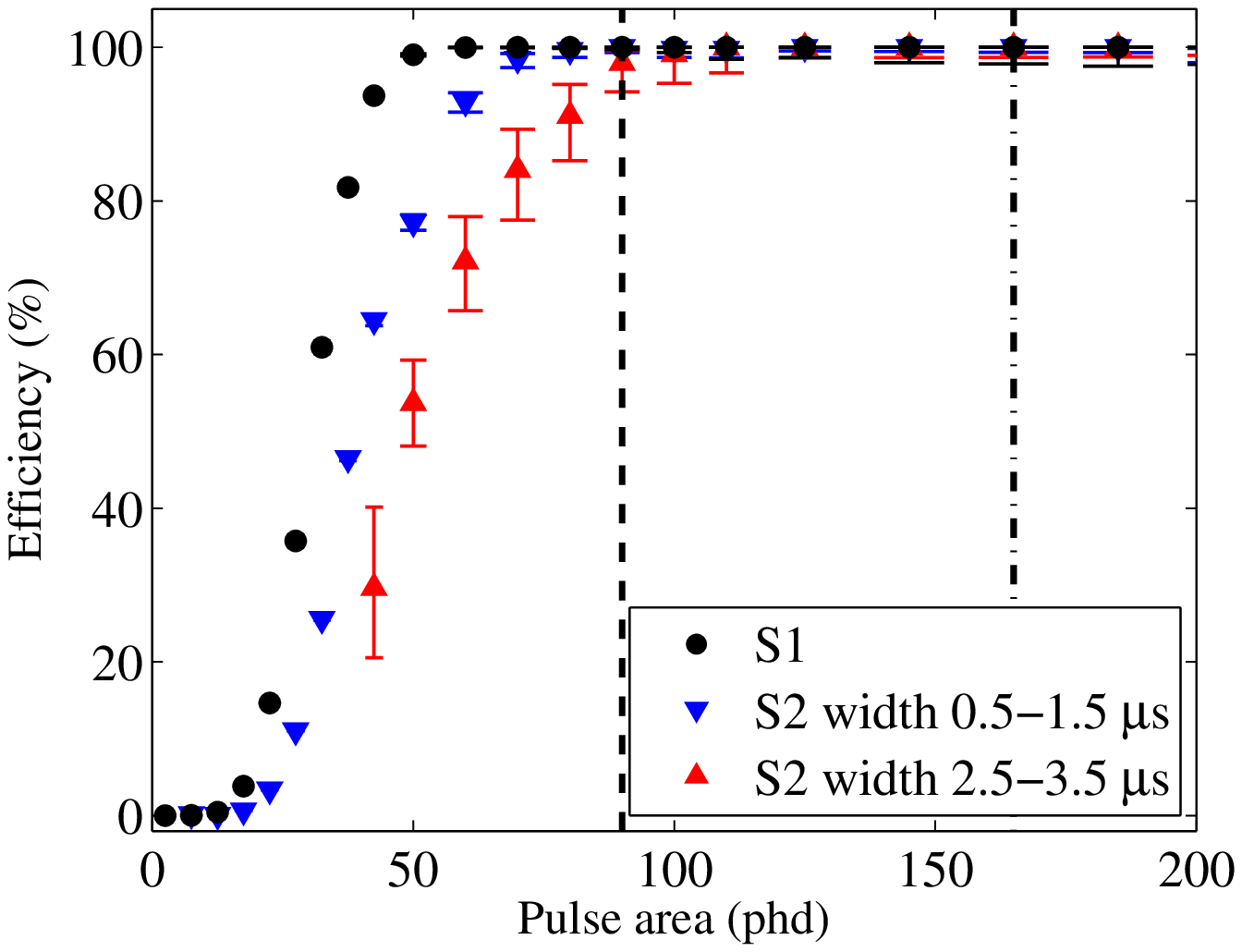}
    \captionof{figure}{Trigger efficiency as function of S2 pulse area for two different width ranges. $\mathcal{E}_{S1}(phd)$ is also shown for comparison. The dashed line shows the pulse area of \mbox{90 phd} at which the efficiencies exceed 98\%. The dash-dotted lines represent the lowest analysis threshold on the S2 of \mbox{165 phd}.}
    \label{Fig:TrgEff_vs_TotalArea_SeparateWidth}
\end{figure}

The effect of pulse width on the trigger efficiency is studied by comparing the pulse width dependence of efficiencies in two S2 pulse area ranges, as shown in Fig. \ref{Fig:TrgEff_vs_Width}. For S2 signals with an area between 50 and \mbox{60 phd}, the trigger efficiency decreases as the pulse width increases beyond \mbox{2 $\mu$s}. The impact of the pulse width is reduced for larger pulse areas, as can be seen in Fig. \ref{Fig:TrgEff_vs_Width} where the efficiency stays above 98\% for all pulse widths when the S2 pulse area is between 100 and \mbox{120 phd}. The efficiency for two S2 pulse width ranges as a function of the total pulse area is shown in Fig. \ref{Fig:TrgEff_vs_TotalArea_SeparateWidth}. This demonstrates the impact on efficiency when the S2 signal is narrower (0.5-\mbox{1.5 $\mu$s}) and wider (2.5-\mbox{3.5 $\mu$s}) than the S2 filter width. For reference, $\mathcal{E}_{S1}(phd)$ is also shown. For pulse areas above \mbox{90 phd}, all trigger efficiencies are above 98\%. Figure \ref{Fig:TrgEff_vs_TotalArea_SeparateWidth} clearly shows that the inefficiency associated with the S2 pulse width does not impact the LUX WIMP search results, which requires the S2 signal to be greater than \mbox{165 phd}. In addition, the comparison between $\mathcal{E}_{S1}(phd)$ and $\mathcal{E}_{S2}(phd)$ for widths between 0.5 and \mbox{1.5 $\mu$s} indicates that the pulse area difference at which the efficiencies start to rise from zero toward 100\%, as seen in Fig. \ref{Fig:TrgEff_vs_TotalArea}, is not due to the pulse width of the S2 signals being larger than the S2 filter width. 

The trigger efficiency fall-off seen in Fig. \ref{Fig:TrgEff_vs_Width} is consistent with the reduction factor of the $Area_{S2F}$ when the pulse is wider than the S2 filter, shown in \mbox{Fig. \ref{Fig:S2F_MeasuredPulseArea_Width}}, and the measured trigger efficiencies, shown in Fig. \ref{Fig:TrgEff_vs_TotalArea_SeparateWidth}. For instance, at a pulse width of \mbox{3 $\mu$s}, the $Area_{S2F}$ is reduced to about 60\% of its actual value. Thus, the $Area_{S2F}$ of the S2 signals with an area of \mbox{55 phd} and width of \mbox{3 $\mu$s} is approximately \mbox{33 phd}. According to Fig. \ref{Fig:TrgEff_vs_TotalArea_SeparateWidth}, $\mathcal{E}_{S2}(phd)$ of the S2s of the width less than \mbox{2 $\mu$s} is about 50\% at \mbox{33 phd}, which is consistent with the measured efficiency shown in Fig. \ref{Fig:TrgEff_vs_Width}. 

\section{Event-based trigger efficiency} \label{section:EventBased_TrgEff}

Some physical quantities are meaningful only when both S1 and S2 signals are considered. Examples are the recoil energy and event location. The trigger efficiencies as a function of these parameters are evaluated for events with valid S1 and S2 signals. A valid S2 signal is required to have a pulse area of at least \mbox{100 phd}. Identical conditions to those used in the LUX WIMP analysis, as discussed in Ref. \cite{LUX_Run3_Reanalysis}, are enforced for each S1-S2 pair to be considered as a valid event. In addition, there must be no other pulses identified as valid S1 or S2 pulses within the window between \mbox{0.5 ms} (about 1.5 times of the maximum drift length) before the S1 signal and \mbox{0.5 ms} after the S2 signal. This requirement ensures that only single scatter events for which the S1 and the S2 signal are paired correctly, are included. The event-based trigger efficiency associated with events with this definition is greater than 98\%, since the pulse-based trigger efficiency is already above 98\% at an S2 pulse area of \mbox{100 phd}. 

\subsection{Dependence on Recoil Energy}
The dependence of the trigger efficiency on recoil energy is studied with two different approaches; by converting the S1 and S2 pulse areas to equivalent recoil energy, and by direct measurement. The conversion approach relies on NEST \cite{NEST_paper_1,NEST_paper_2} to predict expected detected S1 and S2 pulse areas for different recoil energies which are not corrected for detector effects such as liquid xenon purity. This provides a mapping between the recoil energies and the uncorrected pulse areas to convert the pulse-based trigger efficiencies of S1 and S2 from pulse area ($\mathcal{E}_{S1}(phd)$ and $\mathcal{E}_{S2}(phd)$) to equivalent recoil energy. Since the S1 signal always precedes the S2 signal of the same event, a trigger signal will be generated by the S2 signal only if the S1 signal fails to generate a trigger. If the S1 signal generates a trigger, the hold-off time will prevent the S2 signal from generating a duplicate trigger for the same event. Thus, the event-based trigger efficiency in this approach can be defined as,

\begin{equation} \label{equation:CombinedTrgEff}
\mathcal{E}_{Evt}(E) = \mathcal{E}_{S1}(E) + (1-\frac{\mathcal{E}_{S1}(E)}{100}) \times \mathcal{E}_{S2}(E),
\end{equation}

\noindent where $\mathcal{E}_{Evt}(E)$ is the event-based trigger efficiency of an event of recoil energy $E$ and $\mathcal{E}_{S1}(E)$ and $\mathcal{E}_{S2}(E)$ are the pulse-based trigger efficiencies of S1 and S2 corresponding to an event of recoil energy $E$, respectively. The trigger efficiency is studied with this method at the equivalent recoil energy between 0.1 and \mbox{13.0 keV} for both ER and NR. 

The dominant uncertainties in the conversion are the correction for photon loss due to light collection efficiency for the S1 signal and the electron loss due to impurities in the liquid xenon for the S2 signal. The correction factors for these two uncertainties are position-dependent. Thus, they cannot be corrected for in the conversion since the event location, which can only be obtained from evaluating the S1 and S2 signals of the same event, is unknown. The uncertainty associated with the light collection efficiency does not change over time. However, the uncertainty associated with the electron loss due to impurities can change over time, depending on the purity of the liquid xenon. Therefore, the trigger efficiency obtained this way represents only the efficiency during the time period for which the liquid xenon purity correction is used in the conversion. Nevertheless, the efficiency obtained from a period of lower xenon purity can be interpreted as a lower bound of any measurement at higher xenon purity. The efficiency during the LUX WIMP search is the focus of this work, thus, the conversion from pulse area to equivalent recoil energy is done using the correction factor associated with the light collection efficiency and the xenon purity as reported in Ref. \cite{LUX_FirstRunResults_2014}. Two separate assumptions are made to account for these uncertainties. The first is to assume a maximum loss in both S1 and S2 signals and the second is to assume a minimum loss. These two assumptions correspond to the lower and the upper bounds of the trigger efficiency at any given time during the LUX WIMP search.

\begin{figure}[!ht]
	\centering
	\includegraphics[width=88mm]{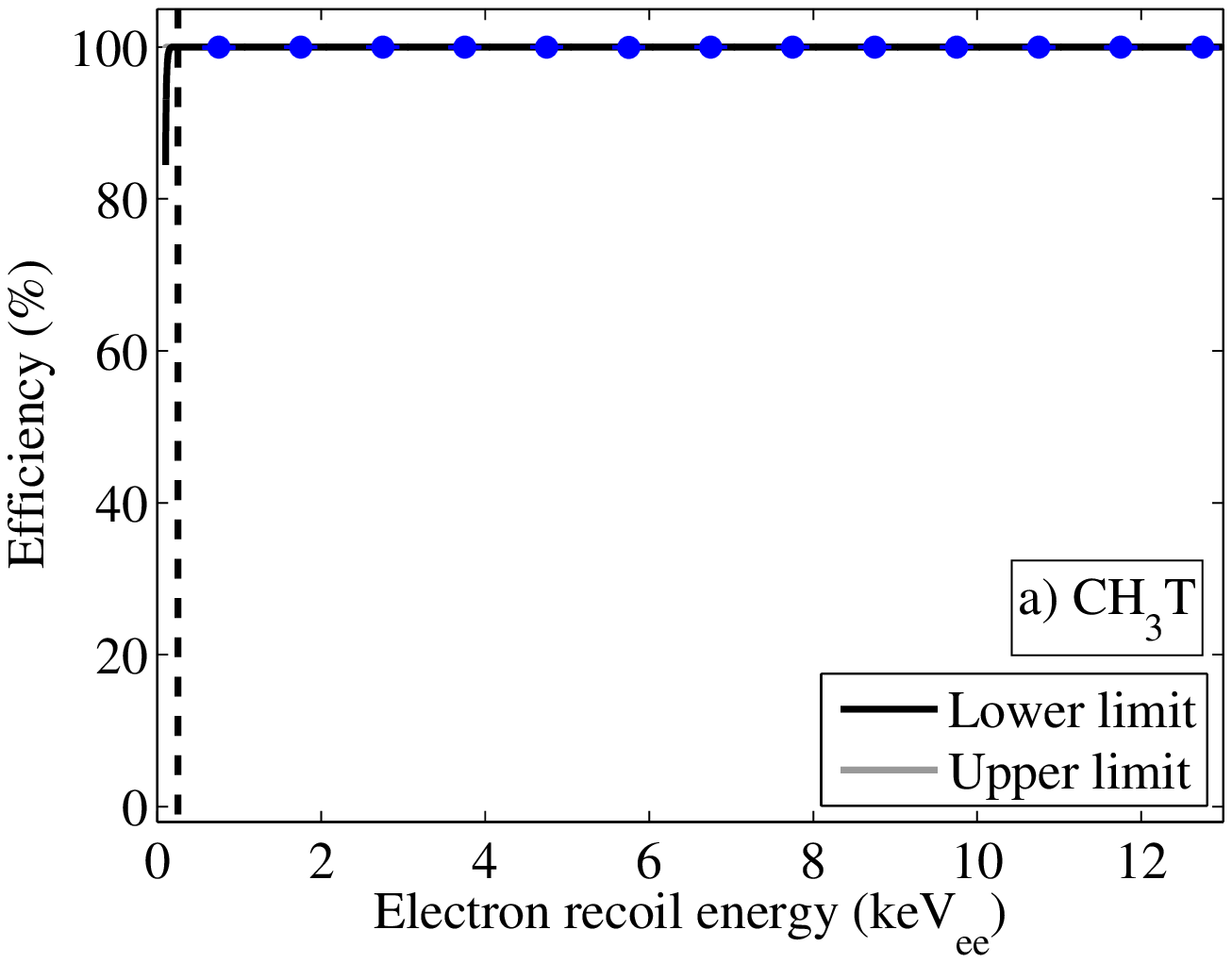}
    \includegraphics[width=88mm]{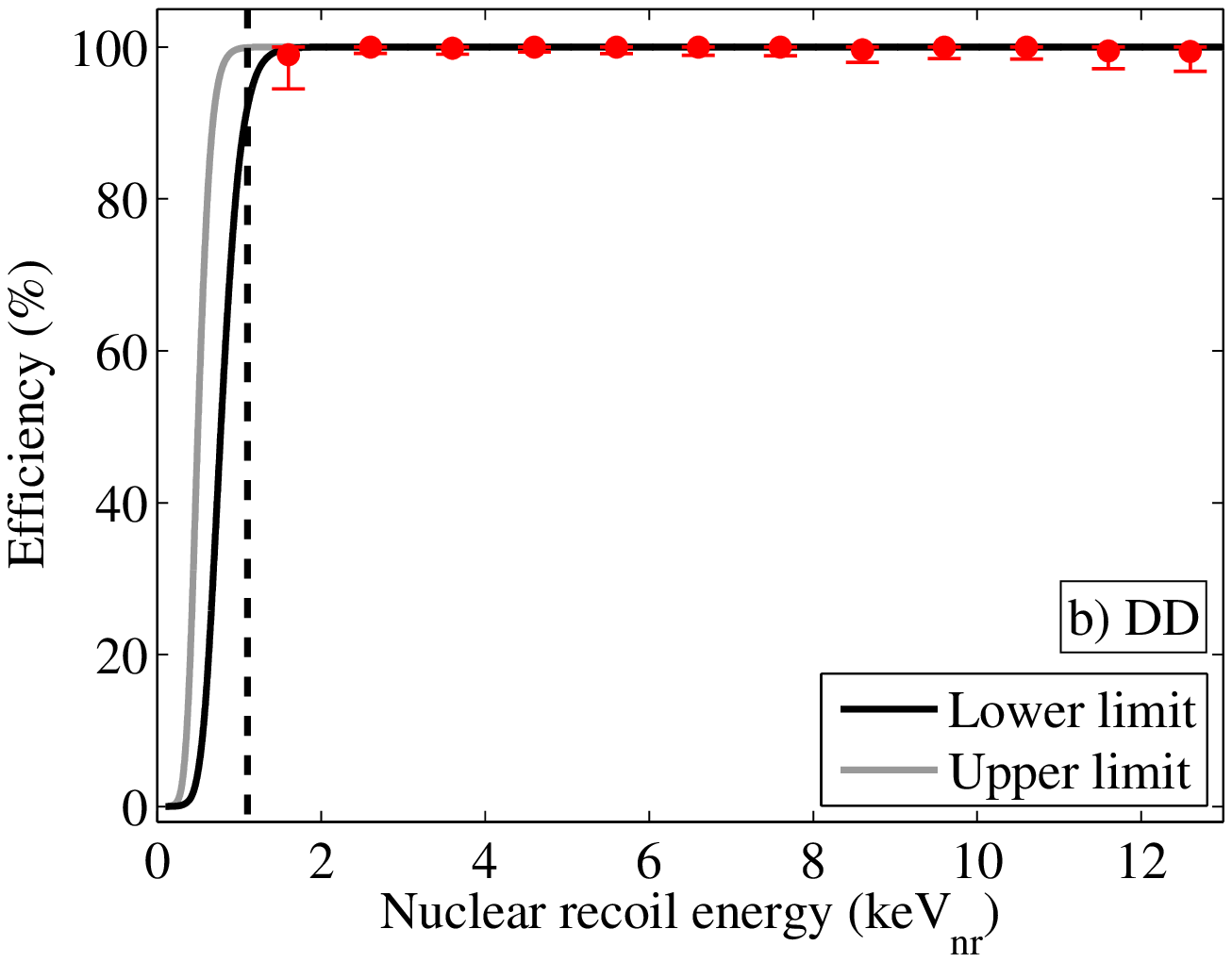}
    \captionof{figure}{Event-based trigger efficiency as a function of reconstructed recoil energy. a) Trigger efficiency for ER. b) Trigger efficiency for NR. In both panels, the black(gray) solid curves indicate the lower(upper) limit of the efficiency from an assumption of having lowest(highest) light collection efficiency and liquid xenon purity level during the LUX WIMP search. They are obtained from converting the pulse area associated with $\mathcal{E}_{S1}(phd)$ and $\mathcal{E}_{S2}(phd)$ to equivalent recoil energy using Eq. \ref{equation:CombinedTrgEff}. The data points are obtained from the direct measurement from CH\textsubscript{3}T and DD calibration data. The vertical dashed lines indicate the lower-end of the bin that contains the event with the lowest energy found in the data sets used in the direct measurement approach.}
    \label{Fig:Evt_TrgEff_vs_Energy}
\end{figure}

In the direct measurement approach, ER and NR events are searched for as described previously. The recoil energy of an ER event is reconstructed according to the relation shown in Eq. \ref{equation:EnergyRecon} \cite{EnergyRecon_g1g2}.

\begin{equation} \label{equation:EnergyRecon}
\frac{E[\mathrm{keV_{ee}}]}{W} = \frac{S1_c}{g_1}+\frac{S2_c}{g_2}
\end{equation}

\noindent $E[\mathrm{keV_{ee}}]$ is the reconstructed energy in ER equivalent scale. The values $g_1$ and $g_2$ are measured to be 0.117 $\pm$ \mbox{0.003 phd per photon} and 12.1 $\pm$ \mbox{0.8 phd per ionization electron}, respectively \cite{LUX_Run3_Reanalysis}. The value of the work function $W$ is taken to be 13.7 $\pm$ \mbox{0.2 keV} \cite{Dahl_thesis_2009_Xenon10_EnergyRecon}. The $S1_c$ and $S2_c$ are position-corrected S1 and S2 pulse areas, respectively. For NR events, the recoil energy is reconstructed according to Lindhard's theory \cite{LindhardTheory}. The recoil energy is reconstructed in an ER equivalent energy scale using Eq. \ref{equation:EnergyRecon}, then converted to an NR equivalent scale using Eq. \ref{equation:LindhardModel}. 

\begin{equation} \label{equation:LindhardModel}
E[\mathrm{keV_{nr}}] = \frac{E[\mathrm{keV_{ee}}]}{L[\mathrm{keV_{nr}}]}
\end{equation}

\noindent $E[\mathrm{keV_{nr}}]$ is the reconstructed energy in NR equivalent scale. The factor $L[\mathrm{keV_{nr}}]$ is the Lindhard's factor, described in Ref. \cite{LindhardTheory}. A detailed discussion of this procedure is described in Ref. \cite{LUX_Run3_Reanalysis,LUX_DD_Paper}. 

Figure \ref{Fig:Evt_TrgEff_vs_Energy} shows the trigger efficiency as a function of the reconstructed recoil energy for both approaches. It was found that the lower bound of the trigger efficiency from the conversion approach exceeds 98\% at a recoil energy of \mbox{0.2 $\mathrm{keV_{ee}}$} and \mbox{1.3 $\mathrm{keV_{nr}}$} for ER and NR, respectively. Due to the \mbox{100 phd} threshold used in defining a valid S2 in the measurement approach, recoil events that produce S2s with an area below \mbox{100 phd} are excluded from the efficiency measurement. The lowest recoil energies that are found in the data included in this analysis are \mbox{0.3 $\mathrm{keV_{ee}}$} and \mbox{1.1 $\mathrm{keV_{nr}}$} for ER and NR events, respectively. The bins used in this analysis start from these two numbers with a bin size of \mbox{1 keV} in both ER and NR. The efficiencies from the conversion approach at these two energies are greater than 99\% for ER and 92\% for NR. For ER, the efficiency from the direct measurement varies between 99\% and 100\% between recoil energies of 0.3 and \mbox{13.0 $\mathrm{keV_{ee}}$}. The average lower uncertainty is 0.1\% with the largest value of 0.2\% at the last energy bin. For NR between 1.1 and \mbox{13.0 $\mathrm{keV_{nr}}$}, the efficiency from the direct measurement varies between 99\% and 100\%. The average lower uncertainty is 1.6\% with the largest value of 4.5\%. A comparison with the efficiency curves obtained from the conversion approach shows excellent agreement.

\subsection{Dependence on S1-S2 time separation}
A reduction in the trigger efficiency associated with the reset time of the LUX trigger system is expected to be observed. This inefficiency occurs when only one trigger group has an S2 filter output greater than the low filter threshold during the trigger search window. Since the coincidence requirement is not met, no trigger signal is generated, and the LUX trigger system is reset to idle state at the end of the search window. The reset process is estimated to take approximately \mbox{4.1 $\mu$s} \cite{Eryk_Thesis}. During this period, the LUX trigger system will not process any input signals.

\begin{figure}[ht]
	\centering
	\includegraphics[width=88mm]{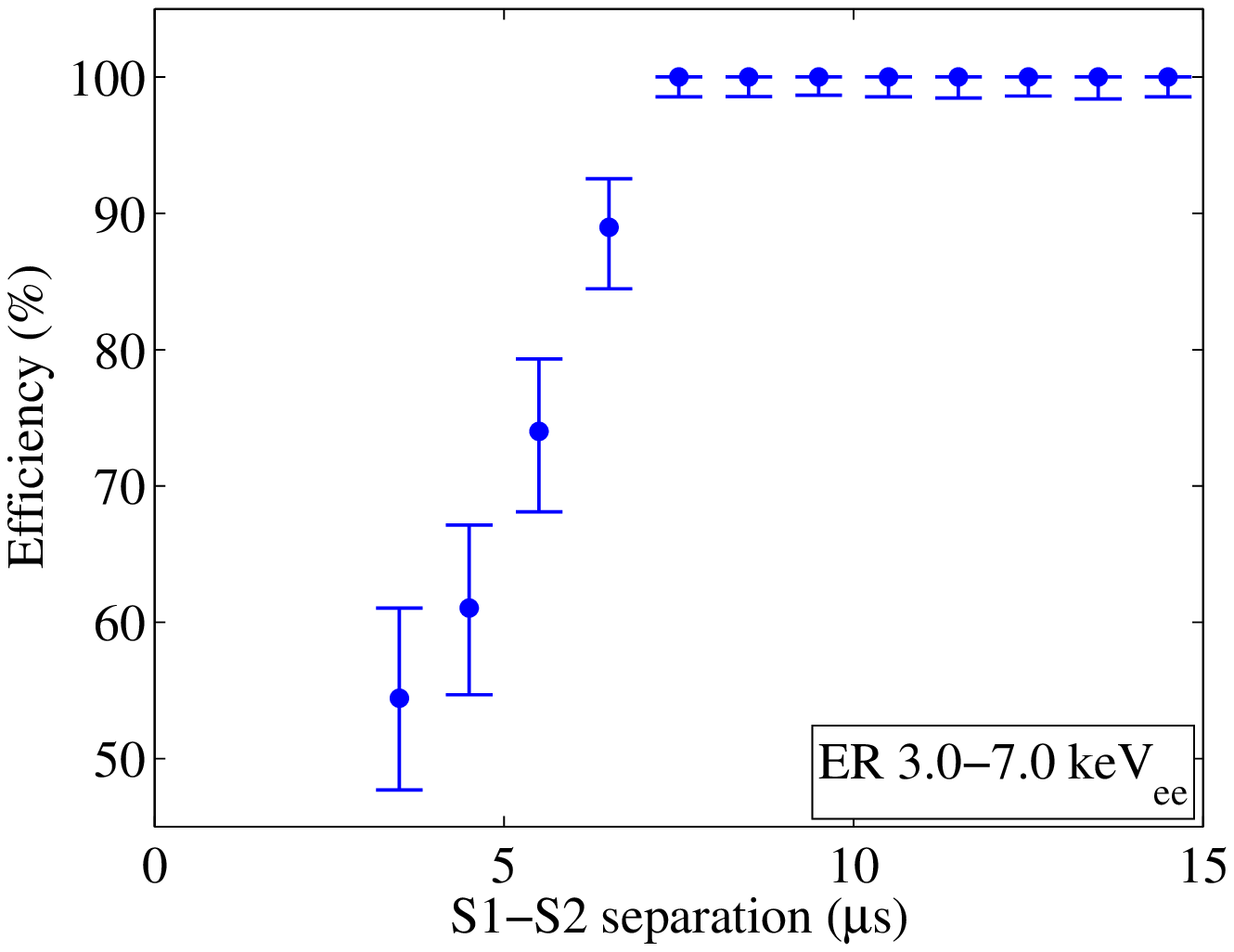}
    \captionof{figure}{Event-based trigger efficiency as a function of the time separation between the start times of the S1 and the S2 signals of ER events from CH\textsubscript{3}T with reconstructed recoil energy between 3.0 and \mbox{7.0 $\mathrm{keV_{ee}}$}.}
    \label{Fig:TrgEff_vs_S1S2Separation}
\end{figure}

To study this effect, the trigger efficiency was measured as a function of time separation between the start times of the S1 and the S2 signals for ER events from CH\textsubscript{3}T. The ER events are chosen because the events from CH\textsubscript{3}T are uniformly distributed inside the detector, and thus have all possible S1-S2 time separations within the maximum drift time, including very short separations. NR events from DD have a fixed drift time due to the location of the neutron beam in the liquid xenon and the S1-S2 separation is not within the time range of interest. The energy range of 3.0-\mbox{7.0 $\mathrm{keV_{ee}}$} is chosen because ER events in this energy range produce S1 signals between 20 and \mbox{50 phd} \cite{LUX_CH3T_Calib} which have trigger efficiencies below 100\%, as shown in Fig. \ref{Fig:TrgEff_vs_TotalArea}. At energies below \mbox{3.0 $\mathrm{keV_{ee}}$}, the S1 signals are too small to start the search window, while at energies above \mbox{7.0 $\mathrm{keV_{ee}}$}, the S1 signals are large enough to generate the trigger. The result of this study is shown in Fig. \ref{Fig:TrgEff_vs_S1S2Separation}. The fall-off at a separation time of around \mbox{7 $\mu$s} agrees very well with the estimate of the trigger reset-time. The main contributions to the \mbox{7 $\mu$s} are \mbox{2 $\mu$s} from the trigger search window, \mbox{4.1 $\mu$s} from the reset time, and approximately \mbox{1 $\mu$s} for an S2 filter output of an S2 signal to cross the threshold, as shown in Fig. \ref{Fig:TrgPulseTimeDist}. The trigger decision process, which is started by the S1 signal, always finishes before the S2 signal if the S1-S2 separation is greater than \mbox{7 $\mu$s}. The S1 and S2 signals in this case are treated independently by the trigger system, except if the S1 signal generates a trigger, the S2 signal will occur within the hold-off time period. However, the effect of the hold-off time does not impact the trigger efficiency. 

\subsection{Dependence on event location}
\begin{sidewaysfigure*} 
    \centering
    \includegraphics[width=1.0\linewidth]{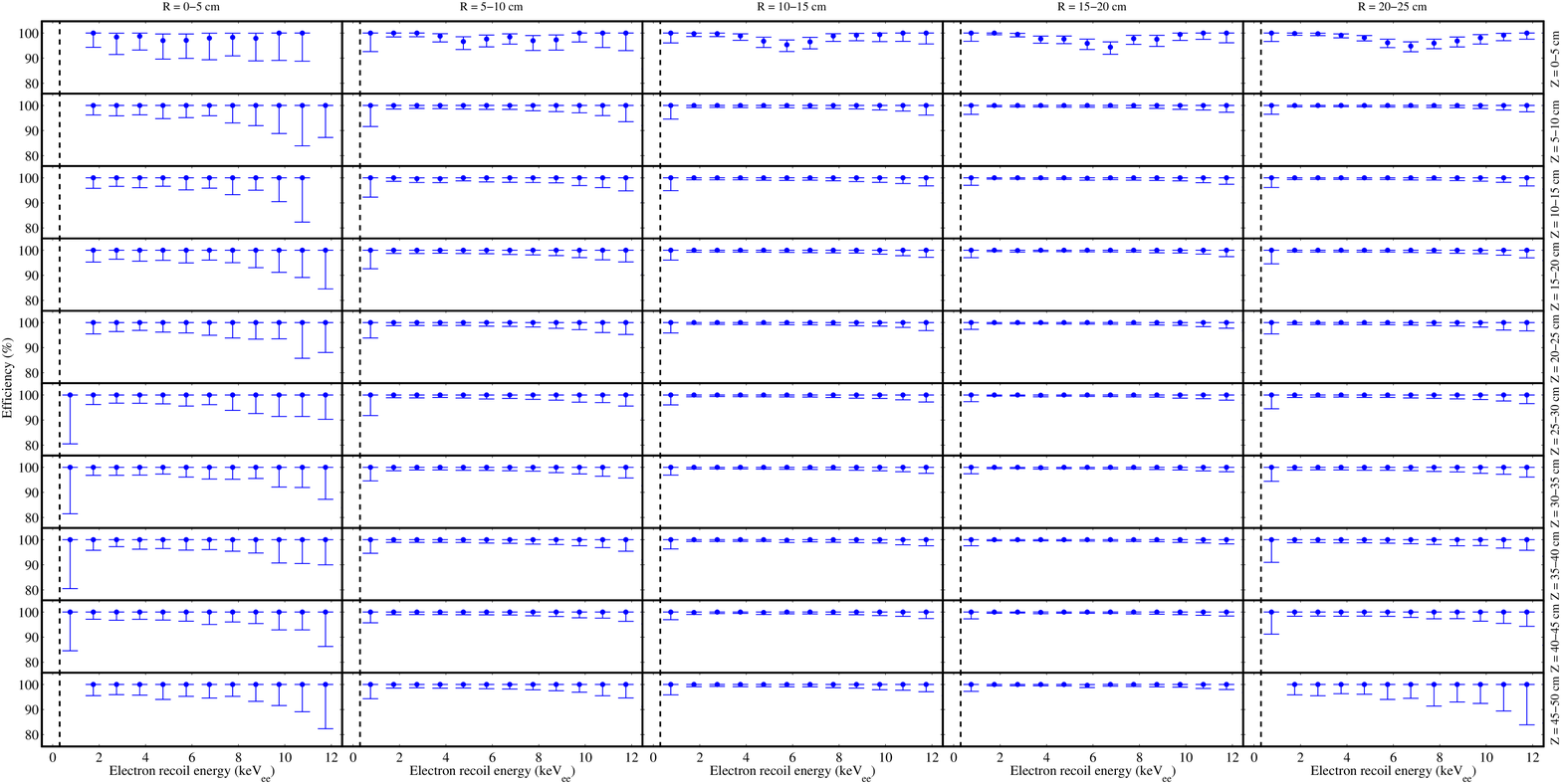}
    \captionof{figure}{The trigger efficiency as a function of the reconstructed recoil energy from the CH\textsubscript{3}T data measured at different positions inside the LUX detector. The panels from left to right within the same row are from the same z-position, but different radial positions from innermost sub-volume to the outermost sub-volume as indicated above the plot. The panels from the top to the bottom of the same column are measured from the same radial position, but different z-positions from the top of the detector to the bottom of the detector as indicated on the right-handed side. The absence of data points and limited statistics in some panels are due to their corresponding sub-volumes are small or close to the boundary of the active region. The vertical dashed lines at \mbox{0.3 $\mathrm{keV_{ee}}$} indicate the lowest recoil energy observed in the data set included.}
    \label{Fig:TrgEff_vs_EnergyCH3T_RZ}
\end{sidewaysfigure*}

Since the LUX detector is large, events that have the same recoil energy, but occur at different positions, can have different characteristics and the trigger efficiency can vary across the detector. This effect is studied using the CH\textsubscript{3}T data by partitioning the liquid xenon region inside the detector into many ring-shaped sub-volumes based on depth (z-position) and radius (r-position). Each sub-volume has a vertical thickness of \mbox{5 cm} and a radial thickness of \mbox{5 cm}. The same LUX WIMP analysis quality cuts previously described are used. The results are shown in Fig. \ref{Fig:TrgEff_vs_EnergyCH3T_RZ}. Apart from the inefficiency found in the topmost sub-volumes, which are outside of the fiducial volume used in the LUX WIMP analysis \cite{LUX_Run3_Reanalysis}, the trigger efficiency is essentially uniform across the detector volume varying between 99\% and 100\%. The average lower uncertainty, which is obtained by averaging the lower uncertainties of all data points from the same sub-volume, varies from 1.1\% to 7.7\%. The striking feature in the topmost rings at all radial positions is directly related to the reset-time as mentioned previously. Events that have S1-S2 separation less than \mbox{7 $\mu$s} occur at less than \mbox{1.3 cm} below the liquid xenon surface and, thus, are included in the topmost sub-volumes.

\section{Summary} \label{section:summary}
The trigger efficiency of the LUX trigger system is fully understood in the entire WIMP search region. Both pulse-based and event-based efficiencies are sufficiently high for WIMP searches. The pulse-based trigger efficiency is measured to be greater than 98\% for both S1 and S2 signal with pulse areas greater than \mbox{90 phd}, which is lower than the analysis threshold used in the LUX WIMP analysis (\mbox{165 phd} for S2 signals). The event-based trigger efficiency obtained from conversion approach indicates that the trigger efficiency exceeds 98\% at the recoil energy of \mbox{0.2 $\mathrm{keV_{ee}}$} and above for ER, and \mbox{1.3 $\mathrm{keV_{nr}}$} and above for NR. The direct measurement approach also shows an excellent agreement. It is confirmed that the trigger efficiency is essentially uniform within the fiducial volume of the liquid xenon for the WIMP searches.

\section{Acknowledgments}
\input{20180130_grants_list}

\bibliographystyle{elsarticle-num-names_2}
\bibliography{reference_short_author_2}

\end{document}

%% file: 20180901-lux-tex-author-list_edited.tex
\author[CWR,SLA,KIP]{D.S.~Akerib}

\author[UWM]{S.~Alsum}

\author[ICL]{H.M.~Ara\'{u}jo}

\author[SDS]{X.~Bai}

\author[UCD]{J.~Balajthy}

\author[SUP]{P.~Beltrame}

\author[UCB]{E.P.~Bernard}

\author[LLN]{A.~Bernstein}

\author[CWR,SLA,KIP]{T.P.~Biesiadzinski}

\author[UCB,LBN,YUD]{E.M.~Boulton}

\author[ULD]{B.~Boxer}

\author[LIP]{P.~Br\'as}

\author[ULD]{S.~Burdin}

\author[USD,SDST]{D.~Byram}

\author[PSU]{M.C.~Carmona-Benitez}

\author[BUD]{C.~Chan}

\author[UCD]{J.E.~Cutter}

\author[SUP]{T.J.R.~Davison}

\author[URD]{E.~Druszkiewicz}

\author[UAS]{S.R.~Fallon}

\author[SLA,KIP]{A.~Fan}

\author[LBN,BUD]{S.~Fiorucci}

\author[BUD]{R.J.~Gaitskell}

\author[UAS]{J.~Genovesi}

\author[DPA]{C.~Ghag}

\author[LBN]{M.G.D.~Gilchriese}

\author[PSU]{E.~Grace}

\author[ULD]{C.~Gwilliam}

\author[UMD]{C.R.~Hall}

\author[UCS]{S.J.~Haselschwardt}

\author[UMA,LBN]{S.A.~Hertel}

\author[UCB]{D.P.~Hogan}

\author[SDST,UCB]{M.~Horn}

\author[BUD]{D.Q.~Huang}

\author[SLA,KIP]{C.M.~Ignarra}

\author[UCB]{R.G.~Jacobsen}

\author[CWR,SLA,KIP]{W.~Ji}

\author[UCB,LBN]{K.~Kamdin}

\author[LLN]{K.~Kazkaz}

\author[URD]{D.~Khaitan}

\author[UMD]{R.~Knoche}

\author[USDP]{E.V.~Korolkova}

\author[LBN]{S.~Kravitz}

\author[USDP]{V.A.~Kudryavtsev}

\author[UCD,LLN]{B.G.~Lenardo}

\author[LBN]{K.T.~Lesko}

\author[BUD]{J.~Liao}

\author[UCB]{J.~Lin}

\author[LIP]{A.~Lindote}

\author[LIP]{M.I.~Lopes}

\author[UCD]{A.~Manalaysay}

\author[TAM,UWM]{R.L.~Mannino}

\author[ICL]{N.~Marangou}

\author[SUP]{M.F.~Marzioni}

\author[UCB,LBN]{D.N.~McKinsey}

\author[USD]{D.-M.~Mei}

\author[URD]{M.~Moongweluwan\corref{cor1}}
\ead{mmoongwe@ur.rochester.edu}

\author[UCD]{J.A.~Morad}

\author[SUP]{A.St.J.~Murphy}

\author[UCS]{C.~Nehrkorn}

\author[UCS]{H.N.~Nelson}

\author[LIP]{F.~Neves}

\author[UCB,LBN]{K.C.~Oliver-Mallory}

\author[UWM]{K.J.~Palladino}

\author[UCB,LBN]{E.K.~Pease}

\author[UAS]{G.R.C.~Rischbieter}

\author[BUD]{C.~Rhyne}

\author[USDP]{P.~Rossiter}

\author[UCS,DPA]{S.~Shaw}

\author[CWR,SLA,KIP]{T.A.~Shutt}

\author[LIP]{C.~Silva}

\author[UCS]{M.~Solmaz}

\author[LIP]{V.N.~Solovov}

\author[LBN]{P.~Sorensen}

\author[ICL]{T.J.~Sumner}

\author[UAS]{M.~Szydagis}

\author[SDST]{D.J.~Taylor}

\author[BUD]{W.C.~Taylor}

\author[YUD]{B.P.~Tennyson}

\author[TAM]{P.A.~Terman}

\author[SDS]{D.R.~Tiedt}

\author[CSU]{W.H.~To}

\author[UCD]{M.~Tripathi}

\author[UCB,LBN,YUD]{L.~Tvrznikova}

\author[DPA]{U.~Utku}

\author[UCD]{S.~Uvarov}

\author[UCB]{V.~Velan}

\author[BUD]{J.R.~Verbus}

\author[TAM]{R.C.~Webb}

\author[TAM]{J.T.~White\fnref{fn1}}

\author[CWR,SLA,KIP]{T.J.~Whitis}

\author[LBN]{M.S.~Witherell}

\author[URD]{F.L.H.~Wolfs}

\author[PSU]{D.~Woodward}

\author[LLN]{J.~Xu}

\author[ICL]{K.~Yazdani}

\author[USD]{C.~Zhang}

\address[CWR]{Case Western Reserve University, Department of Physics, 10900 Euclid Ave, Cleveland, OH 44106, USA}

\address[SLA]{SLAC National Accelerator Laboratory, 2575 Sand Hill Road, Menlo Park, CA 94205, USA}

\address[KIP]{Kavli Institute for Particle Astrophysics and Cosmology, Stanford University, 452 Lomita Mall, Stanford, CA 94309, USA}

\address[UWM]{University of Wisconsin-Madison, Department of Physics, 1150 University Ave., Madison, WI 53706, USA}

\address[ICL]{Imperial College London, High Energy Physics, Blackett Laboratory, London SW7 2BZ, United Kingdom}

\address[SDS]{South Dakota School of Mines and Technology, 501 East St Joseph St., Rapid City, SD 57701, USA}

\address[UCD]{University of California Davis, Department of Physics, One Shields Ave., Davis, CA 95616, USA}

\address[SUP]{SUPA, School of Physics and Astronomy, University of Edinburgh, Edinburgh EH9 3FD, United Kingdom}

\address[UCB]{University of California Berkeley, Department of Physics, Berkeley, CA 94720, USA}

\address[LLN]{Lawrence Livermore National Laboratory, 7000 East Ave., Livermore, CA 94551, USA}

\address[LBN]{Lawrence Berkeley National Laboratory, 1 Cyclotron Rd., Berkeley, CA 94720, USA}

\address[YUD]{Yale University, Department of Physics, 217 Prospect St., New Haven, CT 06511, USA}

\address[ULD]{University of Liverpool, Department of Physics, Liverpool L69 7ZE, UK}

\address[LIP]{LIP-Coimbra, Department of Physics, University of Coimbra, Rua Larga, 3004-516 Coimbra, Portugal}

\address[USD]{University of South Dakota, Department of Physics, 414E Clark St., Vermillion, SD 57069, USA}

\address[SDST]{South Dakota Science and Technology Authority, Sanford Underground Research Facility, Lead, SD 57754, USA}

\address[PSU]{Pennsylvania State University, Department of Physics, 104 Davey Lab, University Park, PA  16802-6300, USA}

\address[BUD]{Brown University, Department of Physics, 182 Hope St., Providence, RI 02912, USA}

\address[URD]{University of Rochester, Department of Physics and Astronomy, Rochester, NY 14627, USA}

\address[UAS]{University at Albany, State University of New York, Department of Physics, 1400 Washington Ave., Albany, NY 12222, USA}

\address[DPA]{Department of Physics and Astronomy, University College London, Gower Street, London WC1E 6BT, United Kingdom}

\address[UMD]{University of Maryland, Department of Physics, College Park, MD 20742, USA}

\address[UCS]{University of California Santa Barbara, Department of Physics, Santa Barbara, CA 93106, USA}

\address[UMA]{University of Massachusetts, Amherst Center for Fundamental Interactions and Department of Physics, Amherst, MA 01003-9337 USA}

\address[USDP]{University of Sheffield, Department of Physics and Astronomy, Sheffield, S3 7RH, United Kingdom}

\address[TAM]{Texas A \& M University, Department of Physics, College Station, TX 77843, USA}

\address[CSU]{California State University Stanislaus, Department of Physics, 1 University Circle, Turlock, CA 95382, USA}

\cortext[cor1]{Corresponding author}
\fntext[fn1]{deceased}

%% file: 20180130_grants_list.tex
This work was partially supported by the U.S. Department of Energy (DOE) under award numbers DE-AC02-05CH11231, DE-AC05-06OR23100, DE-AC52-07NA27344, DE-FG01-91ER40618, DE-FG02-08ER41549, DE-FG02-11ER41738, DE-FG02-91ER40674, DE-FG02-91ER40688, DE-FG02-95ER40917, DE-NA0000979, DE-SC0006605, DE-SC0010010, DE-SC0015535, and DE-SC0017891; the U.S. National Science Foundation under award numbers PHYS-0750671, PHY-0801536, PHY-1003660, PHY-1004661, PHY-1102470, PHY-1312561, PHY-1347449, PHY-1505868, and PHY-1636738; the Research Corporation grant RA0350; the Center for Ultra-low Background Experiments in the Dakotas (CUBED); and the South Dakota School of Mines and Technology (SDSMT). LIP-Coimbra acknowledges funding from Funda\c{c}\~{a}o para a Ci\^{e}ncia e a Tecnologia (FCT) through the project-grant PTDC/FIS-NUC/1525/2014. Imperial College and Brown University thank the UK Royal Society for travel funds under the International Exchange Scheme (IE120804). The UK groups acknowledge institutional support from Imperial College London, University College London and Edinburgh University, and from the Science \& Technology Facilities Council for PhD studentships ST/K502042/1 (AB), ST/K502406/1 (SS) and ST/M503538/1 (KY). The University of Edinburgh is a charitable body, registered in Scotland, with registration number SC005336. 

The \textsuperscript{83}Rb isotope used in this research were supplied by the United States Department of Energy Office of Science by the Isotope Program in the Office of Nuclear Physics.

This research was conducted using computational resources and services at the Center for Computation and Visualization, Brown University.

The collaboration gratefully acknowledge the logistical and technical support and the access to laboratory infrastructure provided to us by the Sanford Underground Research Facility (SURF) and its personnel at Lead, South Dakota. SURF was developed by the South Dakota Science and Technology Authority, with an important philanthropic donation from T. Denny Sanford, and is operated by Lawrence Berkeley National Laboratory for the Department of Energy, Office of High Energy Physics.